\documentclass[conference]{IEEEtran}
\IEEEoverridecommandlockouts
\usepackage{cite}
\usepackage{amsmath,amssymb,amsfonts}
\usepackage{commath}
\usepackage{algorithmic}
\usepackage{graphicx}
\usepackage{textcomp}
\usepackage{xcolor}
\usepackage{multirow}
\usepackage{epstopdf}

\usepackage{tcolorbox}

\usepackage[T1]{fontenc}
\usepackage{enumitem}
\setitemize[1]{itemsep=0pt,partopsep=0pt,parsep=\parskip,topsep=0pt}
\graphicspath{{./Figures/}}
\def\BibTeX{{\rm B\kern-.05em{\sc i\kern-.025em b}\kern-.08em
    T\kern-.1667em\lower.7ex\hbox{E}\kern-.125emX}}
\begin{document}

%\title{SoK: Trending Topics Analysis in Cybersecurity Texts}
%\title{What Do People Concern about Cybersecurity?}
%\subtitle{An Analysis of Trending Topics in Cybersecurity Texts}

%\title{What do People Talk about in Cybersecurity?\\ {\huge An Analysis of Trending Topics in Cybersecurity Texts}}

\title{Analysis of Trending Topics and Text-based Channels of Information Delivery in Cybersecurity}

\author{
    \IEEEauthorblockN{Tingmin Wu\IEEEauthorrefmark{1}\IEEEauthorrefmark{2}, Wanlun Ma\IEEEauthorrefmark{1}, Sheng Wen\IEEEauthorrefmark{1}, Xin Xia\IEEEauthorrefmark{4}, Cecile Paris\IEEEauthorrefmark{2}, Surya Nepal\IEEEauthorrefmark{2}, Yang Xiang\IEEEauthorrefmark{1}}
    \IEEEauthorblockA{\IEEEauthorrefmark{1}Swinburne University of Technology \IEEEauthorrefmark{2}CSIRO's Data61 
    %\IEEEauthorrefmark{3} University of Electronic Science and Technology of China
    \IEEEauthorrefmark{4}Monash University
    }
}
%\author{Anonymous Authors}
%\author{\IEEEauthorblockN{1\textsuperscript{st} Given Name %Surname}
%\IEEEauthorblockA{\textit{dept. name of organization (of %Aff.)} \\
%\textit{name of organization (of Aff.)}\\
%City, Country \\
%email address or ORCID}
%\and
%\IEEEauthorblockN{2\textsuperscript{nd} Given Name Surname}
%\IEEEauthorblockA{\textit{dept. name of organization (of %Aff.)} \\
%\textit{name of organization (of Aff.)}\\
%City, Country \\
%email address or ORCID}
%\and
%\IEEEauthorblockN{3\textsuperscript{rd} Given Name Surname}
%\IEEEauthorblockA{\textit{dept. name of organization (of %Aff.)} \\
%\textit{name of organization (of Aff.)}\\
%City, Country \\
%email address or ORCID}
%\and
%\IEEEauthorblockN{4\textsuperscript{th} Given Name Surname}
%\IEEEauthorblockA{\textit{dept. name of organization (of %Aff.)} \\
%\textit{name of organization (of Aff.)}\\
%City, Country \\
%email address or ORCID}
%\and
%\IEEEauthorblockN{5\textsuperscript{th} Given Name Surname}
%\IEEEauthorblockA{\textit{dept. name of organization (of %Aff.)} \\
%\textit{name of organization (of Aff.)}\\
%City, Country \\
%email address or ORCID}
%\and
%\IEEEauthorblockN{6\textsuperscript{th} Given Name Surname}
%\IEEEauthorblockA{\textit{dept. name of organization (of %Aff.)} \\
%\textit{name of organization (of Aff.)}\\
%City, Country \\
%email address or ORCID}
%}

\maketitle
\pagestyle{plain}

\begin{abstract}
Computer users are generally faced with difficulties in making correct security decisions. While an increasingly fewer number of people %fewer and fewer of them 
are trying or willing to take formal security training, online sources including news, security blogs, and websites are continuously making security knowledge more accessible. 
%Analysis and comparison of the trending topics in the cybersecurity texts from these sources can help identify the regularity of critical security information delivery to facilitate security education, and also gain the insights of how recent security incidents evolve for people in cybersecurity.
Analysis of  cybersecurity texts can provide insights into the trending topics and identify current security issues as well as how cyber attacks evolve over time. These in turn can support researchers and practitioners in predicting and preparing for these attacks. 
%Surya %surya why researchers only; i would rather say practitioners; happy to say researchers and practitioners 
%tingmin -  added "and practitioners"
Comparing different sources may facilitate the learning process for normal users by persisting the security knowledge gained from different cybersecurity context.
%surya - not clear what do you mean by persistence in knowledge understanding; it needs to be very clear what you mean
%tingmin - revised
Prior studies neither  systematically analysed the wide-range of digital sources nor provided any standardisation in analysing the trending topics from recent security texts. Although LDA has been widely adopted in topic generation, its generated topics cannot cover the cybersecurity concepts completely and considerably overlap. 
%surya - why it does not perform well. We need a sentence to provide a reason. 
%tingmin - revised
To address this issue, we propose a semi-automated classification method to generate comprehensive security categories instead of LDA-generated topics. We further compare the identified 16 security categories across different sources based on their popularity and impact. We have revealed several surprising findings. 
%(1) \textit{information privacy}, \textit{password/encryption}, \textit{mobile/application security} and \textit{network attack} occur frequently in our dataset and have a strong correlation between each other. (2) \textit{cybercriminal activity} has the largest popularity among 16 categories across all the sources. 
(1) The %absolute 
impact reflected from cybersecurity texts strongly correlates with the monetary loss caused by cybercrimes. (2) For most categories, security blogs share the largest popularity and largest absolute/relative impact over time. (3) Websites deliver security information without caring about timeliness much, where one third of the articles do not specify the date and the rest have a time lag in posting emerging security issues.
%The analyses of security issues and differences between the sources 
%Our analyses help pinpoint the major challenges faced by researchers and practitioners and also generate the patterns in delivering security knowledge to normal users.

\end{abstract}


\section{Introduction}
Humans are playing an indispensable role in cybersecurity, and, because of that, are especially targeted in cyber attacks \cite{evans2016human,hadlington2017human}.  CybSafe analysis of UK ICO reports that 90\% of data breaches were caused by human mistakes in 2019 \cite{cybsafe}. Computer users  generally have  difficulties in making security decisions due to lack of knowledge, cognitive limitation or deviations from rationality \cite{acquisti2005privacy}. However, they have to deal with sophisticated intrusions when their security software, such as antivirus or firewall, become obsolete \cite{wash2010folk}. %With the aim to protect users from attackers, some studies generated the model 
To keep users in the loop is vital as any security measures can leave users more vulnerable when they lose resistance to unknown attacks.

End users are expected to learn more about cyber attacks, security measures and the key techniques that keeps them informed  about cyber risks, and they need to take timely actions. Domain knowledge, especially in the cybersecurity field, is not easily turned into cognitive abilities without proper training \cite{ben2015effects}. Nonetheless, formal security education or training is time-consuming and requires users' undivided attention, and one-size-fits-all trainings hardly keep people engaged as they might have different learning preferences or background knowledge \cite{security_training_shortages}. Such unthoughtful schemes can even cause market losses \cite{redmiles2018dancing}. Compared to certification programs, online cybersecurity texts give internet users easier access to  security knowledge in order to make correct decisions in the time of cyber incidents. 

We identified three sources that users often find for security texts from in their daily life:  news, security blogs and websites. %News and websites were also studied in \cite{rader2015identifying}. 
News are published by news agencies as the leading media for general audience. Examples include BBC, USA Today, etc. Security blogs can be more tailored towards security experts or individuals (including general users) who are interested in cybersecurity. These blogs mainly post security articles consisting of the latest threats, experts' opinions and security solutions for both businesses and individuals to use in practice. 
%Websites refer to the security advice (e.g. tips and alerts) and news reported by authorised organisations in government, research or industry to protect their audience. 
Websites include any information provided by authorised organisations (government, research institutes or industries), for the purpose of guiding the readers to behave securely online. 
These sources provide a range of educational materials which can benefit different communities.

The majority of existing analyses have failed to consider all the user-accessible resources in order to provide users with a large selection for informal security learning. This selection could include studies on cyber threats \cite{jang2014survey,tripathi2011taxonomic,massacci2010right,burger2014taxonomy} and threat intelligence \cite{tounsi2018survey,abu2018cyber}.
Several studies \cite{rader2015identifying,sauerwein2019analysis,shillair2017multiple} 
%surya - this citation is required; otherwise, the sentence reads odd. 
%tingmin - added the citations
analysed the security knowledge from multiple sources, but the results are outdated now.  And the data collection was done in a relatively short period (e.g.  from 2011 to 2015 in \cite{rader2015identifying}.), 
Additionally, the trendlines of different topics show how they develop and give direction to ongoing studies, but they have barely been analysed before \cite{sauerwein2019analysis,shillair2017multiple}.
%surya - the trendline does not show the ongoing research; it shows the current threats or hot topics in the community; the research may not align with what is discussed on the news or media; the research is driven by different factors; it may be okay to say industry, but not research. 
%tingmin -  changed 'research' to 'studies'
Some prior research focused on producing security information, but their inferences from  information were biased due to lack of timeliness, or were hard to be adopted in the real world \cite{skopik2016problem,blythe2019security,reeder2017152,redmiles2018first}.
%introduced biases due to only using single source of information or the time lag.
% surya - cite those research 
%tingmin - added the citations
For example, security information sharing informally produces security incident reports, mainly from websites. However, the release requires time to verify whether they meet various standards or not, and might miss the timing of reporting emerging attacks such as zero-day vulnerability \cite{skopik2016problem}. Lack of standardisation also hinders the exchange of security information. %Vendors of IoT devices hardly demonstrate security features or provide handy security advice even in their manuals and support pages \cite{blythe2019security}. 
Moreover, current security advice is usually too technical to understand or not actionable due to their restrictions (e.g., ``never click on links in emails'') \cite{reeder2017152,redmiles2018first}.
%surya I am not sure how the above couple of sentences on standardisation and security advice fit with the rest; you need to fix this. 
%tingmin - revised
Prior works used LDA (Latent Dirichlet Allocation) to cluster security questions \cite{yang2016security,rader2015identifying,chambers2018detecting}. However, traditional LDA does not perform well in capturing domain-specific concepts \cite{park2015efficient}. %because of not dealing with synonymy  
The topics it generates have low granularity and are hard to distinguish.

To address the issue, we propose a semi-automated classification method to generate broad topics in cybersecurity instead of using LDA-generated topics. We first divide our collected security texts into five datasets according to their sources. For each dataset, we run LDA separately. We find that the generated topics by LDA do not capture the domain-specific concepts and are not distinct to each other. To derive more meaningful results, we use the term extraction method to generate a set of terms that summarise the categories for each LDA-generated topic. %As is reported, 
We identify 16 security categories which summarise all those terms. We analyse the popularity and impact of those categories to analyse different cybersecurity trends. More specifically, we compare how the security issues evolve across categories and sources over the last decade. This sheds light into the development of security issues in the past ten years and reveals how challenges emerge, which in turn can be used in the prediction of unknown threats. The analyses of security issues and differences between the sources also generate patterns in delivering security knowledge to the general public.

Our research focuses on answering three research questions:

\noindent\textbf{RQ1.} What are the security issues reported in cybersecurity texts?

We discovered 16 security categories for cybersecurity texts from news, security blogs and websites. They can summarise  security issues, including the types of cyber attacks and security techniques. We found that information privacy still remained a dominant topic in the last decade, and this was largely due to criminal offence (including password attack), mobile application attack, and network attack. We also noticed that most articles (83\%) discussed multiple security issues (relevant to up to six security categories).
%\noindent\textbf{RQ2.} How do the security topics fit into or move across the security categories?

\noindent\textbf{RQ2.} How have the security categories varied and evolved over the last decade?

%The categories \textit{information privacy}, \textit{security software/service}, and \textit{cybersecurity program} share the largest popularity, at around 40\%.
Cybercriminal activity has been the most popular and was discussed in most security articles (65\%), followed by the privacy issue, preventive measures (i.e. cybersecurity software, service, and program), with similar popularities at 40\%;
The increase of the absolute impact of the security categories indicates security incidents evolution in both amount and sophistication over the last decade. Security issues in mobile/application and information privacy gained the largest absolute/relative impact over time. The explosion of ransomware (e.g. WannaCry) brought the absolute impact of \textit{malware/virus} to its peak and exceeded the values of all the other categories. The overall absolute impact of all the security categories strongly correlates with the economic loss caused by cybercrimes. Election security has gained a sudden increase in the absolute impact in 2016, which coincides with the U.S. presidential election campaign.

\noindent\textbf{RQ3.} How have the security categories varied and evolved across different sources on cybersecurity over the last decade?

Almost all the categories are popularly present within the three sources, except the categories \textit{election security} and \textit{false/misleading claim} that are only prevalent in news and webs, respectively. Among the three sources, security blogs have largest popularity and absolute/relative impact over time in the majority of categories. The absolute impact of news and security blogs shows upward trends for all the categories in the 2010s, while \textit{false/misleading claim} has a downtrend in webs. Security issues in mobile/application have been the most influential in news and security blogs during the past ten years, followed by the privacy issue. %\textit{information privacy}. \textit{IoT threat}
Threats in IoT show comparable absolute/relative impact to the privacy issue in news. News and security blogs report security events for the first time at similar speed on most categories. Websites deliver security information without caring much about timeliness, with one third of the articles not specifying the date and the rest having a time lag in posting emerging security issues.

We list our contributions as follows:

\begin{itemize}[leftmargin=*]
\item We build a large collection of cybersecurity texts (187,319 articles) from three online sources: news, security blogs, and websites. 
\item We propose a semi-automated classification with combining the term extraction method and the open card sorting to derive the categories of our texts instead of using LDA-generated topics. We identify 16 security categories to analyse the security issues.
\item We conduct an empirical study to analyse the comparison and evolution of the security categories over the last decade as well as across the sources to shed light on the trends of security issues. 
\end{itemize}
%surya - where is the list of contributions?
%tingmin - added the contributions
The rest of our paper is organised as follows. Section~\ref{s_related_work} reviews the related work. Section~\ref{s_research_setting} introduces our research questions and methodology. We present our findings and results in Section~\ref{s_result}. Section~\ref{s_discussion} discusses the limitations of our work. We make conclusions and propose the scope of future work in Section~\ref{s_conclusion}.

\section{Related Work}
\label{s_related_work}

In this section, we review the existing works on users' selection of security information sources, security decision making and learning about security.

%\subsection{Selection of security information sources}
\textbf{Selection of security information sources}.
There is a proven relationship  between security information sources and users' online experience about security and privacy \cite{redmiles2017digital,redmiles2016learned}.  Users are different in their engagement in security protection scenarios and, thus, have different demands of expertise from the sources \cite{forget2016or}.
Redmiles et al. measured the readability of online security advice and examined its correlation with user-reported understanding \cite{redmiles2018first}. Reader and Wash identified patterns from informal sources of security information that help users seek useful data in order to behave safely or solve potential risks \cite{rader2015identifying}. Ion et al. compared the security practices from different people. They found that experts mainly suggest regarding security updates and using password managers \cite{ion2015no}. In contrast, non-experts mostly suggest clicking only on official websites links and regular password changes. Sauerwein1a et al. compared public sources of security information \cite{sauerwein2019analysis}. Das et al. studied how different users gain different information from security news \cite{das2018breaking}. Shillair and Meng compared the impact of different sources in changing users' security behaviours \cite{shillair2017multiple}. Acar et al. analysed informal sources that  provide security guidance in coding for software developers \cite{acar2017developers}. Tounsi and Rais extracted a taxonomy of current threat intelligence types from the literature \cite{tounsi2018survey}.

Users often seek security information from multiple sources while  considering different factors. Nthala and Flechais performed qualitative studies and showed that people applied some measures such as professional level, academic standing and negative experience of the sources \cite{nthala2018informal}. Redmiles et al. found that users measure the trustworthiness of cybersecurity information by the sources for digital security advice and by the content for physical security advice \cite{redmiles2016think}. Nicholson et al. conducted interviews with elderly people about their choices for cybersecurity-related sources of information  and found that they preferred social sources over experts' advice \cite{nicholson2019if}.

%\subsection{Security decision making}
\textbf{Security decision making}.
Wash studied home computer users' understanding of security and built eight folk models to help them make  correct decisions \cite{wash2010folk}. Redmiles et al. exposed that end users  are likely to make optimal decisions when are faced with higher risks \cite{redmiles2018dancing}. Faklaris et al. made a scale to measure users' security-related attitudes \cite{faklaris2019self}. The impact of sources on code security was also studied in \cite{acar2016you}. In addition, even technically savvy people are challenged in making security decisions. Krombholz et al. studied the issue of weak TLS (Transport Layer Security) configurations and found that most of the interviewed administrators had difficulties in deploying HTTPS securely \cite{krombholz2017have}. 

Many factors can affect users' decisions, such as the cost of protection, including time, money and the amount of effort required \cite{nthala2017if}. Howe et al. found users might underestimate the chance to be hacked and, as a result, not adopt the security advice \cite{howe2012psychology}. Wash and Reader also pointed out that some adults did not take actions because they had weak beliefs about hackers or viruses \cite{wash2015too}. Similarly, Zou et al. collected a set of users' perception towards the Equifax breach and found that they tended to undervalue the chance to become victims and procrastinated taking actions even though they recognised the risks \cite{zou2018ve}. Later they suggested improving  data breach notices in terms of readability, media penetration, format and risk indication \cite{zou2019beyond,zou2019youmight}. The tendency to apply security updates was also studied in \cite{redmiles2018asking}. Redmiles et al. researched the influence of social community experiences such as Facebook groups \cite{redmiles2019just}. Mathur and Chetty found that most users rejected automatic mobile app updates due to their unpleasant experience in the past \cite{mathur2017impact}. Sawaya et al. explained the differences in security behaviours and security knowledge of people with the difference in culture \cite{sawaya2017self}. Fagan and Khan studied users' considerations on benefit and risk when they decide whether to follow the security advice or not \cite{fagan2018follow,fagan2016they}. The survey conducted by Acquisti et al. highlighted that both  cognitive hurdles and security tools designed by researchers could influence users' choices \cite{acquisti2017nudges}. 

%\subsection{Security learning}
\textbf{Security learning}.
Security education or training has attracted a large amount of research. This includes the studies on  phishing attack prevention \cite{petelka2019put,wen2019hack,hu2018end}, browser warnings \cite{akhawe2013alice,sunshine2009crying} and password protection \cite{fujita2015attempt,schechter2015learning,fennell1stories}. Safa and Solms shed light on how security knowledge can help reduce the risks of cyber incidents \cite{safa2016information}. Some users suffered because they overestimated their  knowledge of security \cite{crossler2017mobile}.
Abu et al. developed  mental models for users to help them protect their privacy, e.g. with E2E encryption \cite{abu2018exploring}. Stevens et al. did threat modelling in different enterprise scenarios and showed its efficacy in security defence \cite{stevens2018battle}. 
Chen et al. \cite{chen2019self}, similar to \cite{trickel2017shell}, designed a desktop game to teach a series of security practices that users can apply in the real world. Golla et al. studied users' understanding of security warnings and designed password-reuse notifications based on their perceptions \cite{golla2018site}. 

Home computer users also adopt security practices from social learning \cite{nthala2018informal,nthala2018rethinking}. Wash and Cooper found that security training based on facts and advice is more effective by experts than by peers and it can help users behave securely online \cite{wash2018provides}. Das et al. emphasised the importance of social influence in users' change of security behaviours and perceptions \cite{das2014increasing,das2014effect}. Rader et al. also indicated that some users learn security incidents from family and friends \cite{rader2012stories}. Hayes et al. studied how people with visual impairments learn security and privacy from their families, friends, experts or  %other 
disabled %individuals
people \cite{hayes2019cooperative}. 

%surya - at the end of the related work you should have a small paragraphs to link the existing works to our work. What makes our works different to them? You need to justify the need of this work in the context of existing literature.  
%tingmin -  added a paragraph to justify the need of our work
While most of the existing works focused on modelling users' security behaviours or developing the tools for security education, there is no recent study analysing the trends of security topics on a large-scale easily accessible online texts. Compared to formal security training, the online resources about cybersecurity (e.g. newspapers, personal stories, online forums, professional guidelines) are more approachable and diverse to seek help and read regularly. The resources resources deliver important information somehow enable users to make good security decisions. We need an empirical study to exhaust the cybersecurity texts and understand what security issues they report and how they evolve over time as well as difference between the sources. Such a study can help improve informal security learning for end users and forecast innovative cyber attacks.

\section{Study Setup}
\label{s_research_setting}
We explain our three research questions and research methodology in detail.
\subsection{Research Questions}
\noindent\textbf{RQ1.} What are the security issues reported in security texts?

Security texts deliver news and articles about cybersecurity for a range of technology enthusiasts and general users. 
%surya - also the general users. 
%tingmin - revised.
They explain the attack techniques and distribute security tips, guidelines and advice for both businesses and home computer users. 
%Systematical discovering and studying the security issues help identify the challenges against cyber attacks which might catch the interests of researchers and practitioners. 
Categorising the security texts can help identify the security issues. The analysis of the issues sheds light on the challenges faced by researchers and practitioners to advance the development of  threat intelligence to protect the security and privacy of online users. In addition, the security issues identification caters the needs and concerns of normal users when seeking security advice online.

%\textbf{RQ2.} How do the security topics evolve in the past 15 years?

\noindent\textbf{RQ2.} How have the security categories varied and evolved over the last decade?

It is critical to update the topic analysis with the most recent posts. Although there are similar works that have studied security topics, their results are not useful anymore since they have been outdated by at least five years. The worldwide financial loss caused by cybercriminals are predicted to be \$6 billion per year in 2021, increasing from \$3 billion in 2015 \cite{cyberatkincrease}. Intrusions become more sophisticated and hackers employ more advanced techniques.  In a recent example, Florida City suffered a ransomware attack in 2019 and had to pay hackers \$600,000 in Bitcoin, a cybercurrency  known for its (partial) anonymity feature.

In the analysis of the latest trends and drawing a big picture for the security issues, we are the first to identify the security categories systematically beyond using LDA. By doing research on the differences and similarities between LDA-generated topics and our defined security categories, we provide more distinct security topics with a more comprehensive analysis.

\noindent\textbf{RQ3.} How have the security categories varied and evolved across different sources on cybersecurity over the last decade?

Different sources can deliver security information in different ways. News articles are generally published by authorised newspapers and report the latest security events. Security blogs also report the latest news on cybersecurity but might give more insights into the key techniques used from research or technical papers. Websites mainly come from organisations such as universities and banks. They commonly focus on providing informal security information such as security advice for educational purposes.

Understanding different topics from distinct sources can help us cater to the needs of users with different backgrounds. Users might also be concerned about different attacks or data breaches to various degrees. For example, employees of tech companies care about  data breaches to comply with the company reputation under the legislation. Different sources have different preferences over featured articles and techniques. How the topics evolve across different sources can help in informing users where to acquire sufficient security knowledge from and in detecting the emerging trends over platforms.

\subsection{Research Methodology}
To answer the three research questions, we collected real-world media texts and conducted comparative analyses which could informally provide end users with security knowledge. We focused on the topic trends across different security categories as well as  different sources to provide insights into how security issues evolve.

\subsubsection{Data Collection}
We collected our cybersecurity texts from three types of sources: news, security blogs, and websites. We mainly focused on the easily accessible online articles which computer users  read to gain security knowledge. We only collected  articles from the year 2000 to the date of paper writing.

We developed a crawler in Python by leveraging Beautiful Soup \cite{richardson2007beautiful} library. We only extracted text contents (including titles) for topic analysis. We stripped out images, videos, and meaningless contents (e.g. navigation menu and contact information). The publication dates of the articles were extracted from the search results or taken out of the text contents for trend analysis.

We selected the sources based on their popularity, impact and relevance. In the following, we explain how we selected the articles and search results from each source.

%We selected the sources based on their popularity and impact as explained in the following.

\textbf{News}. We selected the newspapers published in English with top circulation ($>100,000$) worldwide (e.g. in US, Australia and India). We included all the 16 news sources used in a similar study \cite{rader2015identifying}. In addition, we added three more news sources which have become more prevalent recently, e.g. Herald Sun (circulation: 303,140 in 2018), and Tech News World (Reader purchase $>$\$100 billion per year).

To identify the contents on cybersecurity, we applied 27 keywords as filters to search for relevant articles only. 
% surya - why did you consider those 27 keywords? How do you select them? You need to explain. 
% tingmin - explained in the following sentence.
%In addition to
We included the 25 terms used in \cite{rader2015identifying}, and added two new keywords (`cybersecurity' and `cyber attack') in the set. According to Google Trends data, people have searched the keyword `cybersecurity' seven times more frequently in the past few years. We manually went through the found articles though most of them were applicable.

During collection, four news sources were removed because they restricted reading the articles and required subscription or purchasing membership plans, e.g. The Globe and Mail. We combined the search results of all the keywords and removed the duplicates. Altogether, 68,066 articles were collected from 15 newspapers.

\textbf{Security blogs}. We then collected texts from the blogs on cybersecurity which provided the latest security news or articles for computer users with various levels of tech knowledge. The blogs can feature threat intelligence to educate their audience in taking protective measures against cyber attacks. We selected the blogs according to their popularity in social media as well as the number of times recommended by Google (e.g., The Hacker News have more than 2 million followers on Facebook). We also included the blogs used in \cite{liao2016acing}. 41 blogs remained after removing the ones with non-text posts, such as commands, attached files and images. We also manually verified the contents to confirm their relevance to cybersecurity. In total, we collected 109,587 articles from the blogs.

\textbf{Webs}. %We followed the venue for collecting the news.
We extended the domain of web pages used in the existing studies to cover all the applicable ones.
A similar study \cite{rader2015identifying} collected the web pages with which  organisations delivered information or instructions on cybersecurity to their employees in order to help them be aware of risks and behave safely online. We classified the organisations that  provide this information into three types: governmental (federal/state government agencies), industrial (telecommunications companies, social network companies and banks) and academic (universities and research agencies). In addition to the web pages used in \cite{rader2015identifying}, we collected more pages from the top-ranked organisations in our country and divided them into the above-mentioned classes.

We applied the 45 keywords used for web page search in a study \cite{rader2015identifying}, combined the search results and removed the duplicates. We also removed the ones that were empty or not security-related. The final collected dataset contained 41,394 articles from 41 webs (17 governmental,  15 industrial and 9 academic).
%We identy the search results, we 
%\textbf{Analysis}.

\subsubsection{Generation of Topics}
To identify what is being discussed in cybersecurity texts, we applied a topic modelling algorithm to extract/generate  topics from our collected articles. We employed LDA (Latent Dirichlet allocation) as 
%it has demonstrated great capability in capturing security topics 
it was used in similar prior studies \cite{rader2015identifying,mei2015security,zaman2011security,neuhaus2010security}. LDA is a probabilistic model that for each document,  gives a set of topic probabilities. Each topic is a set of words with different weights \cite{blei2003latent}. The model considers word occurrences and co-occurrences within a document as well as across different documents in the whole corpus.

We used LDA to extract topics from the security texts. We formed five datasets from our three sources, as shown in Table \ref{t_corpus_analysis}. We ran LDA on each dataset separately,  since LDA is proven to be biased with large datasets  \cite{hu2015modeling}. We implemented the algorithm in Python by using its 'gensim' library \cite{rehurek_lrec}. We identified the optimal number of topics based on topic coherence \cite{roder2015exploring}. The results are depicted in Table \ref{t_corpus_analysis}, and optimal number selection is detailed in Appendix \ref{a_LDA_optimal_number_selection}.

\begin{table}[t]
\centering
\caption{Articles statistics per dataset after sanitisation.} %Number of articles, number of topics and number of words per article.}
\label{t_corpus_analysis}
\begin{tabular}{lrrrr}
\hline
\multirow{2}{*}{Dataset} & \multirow{2}{*}{\#Articles} &\multirow{2}{*}{\#Topics} & \multicolumn{2}{c}{Article length} \\
\cline{4-5}
    &  & & Mean & SD \\
\hline
 News & 51,685 & 18 & 822 & 873  \\
 Security blogs & 108,354 & 15 & 906 & 2,248  \\
 Webs/Governmental & 9,618 & 15 & 655 & 1,533 \\
 Webs/Industrial & 16,810 & 10 & 716 & 1,475  \\
 Webs/Academic & 852 & 10 & 813 & 1,514  \\
\hline
\end{tabular}
\end{table}

%Fig.~\ref{f_article_no_trend} shows the trend of published articles in the last 20 years. Most cybersecurity texts came to surface after 2010, with regular posting afterwards. One can see that security blogs account for the majority of the online resources on security. With around half of the security blogs in post numbers, news volume shows a dramatic increase in the last year. Websites form a relative small portion, with a gentle growth. Two peak points (during 2016 and 2018) can be identified in industrial websites, and it has also jumped since last year. We observed that the number of published articles per month in the 2000s is far smaller than the number in the later ten years. Therefore, we mainly focused on the analysis of trends in the 2010s in our study.

\textbf{Articles sanitisation}. Based on the generated topics, we removed the articles which were not related to security. As each generated topic was presented as a list of words, we inferred the conceptually specific topics by reading and understanding the combinations. We selected the topics whose all words were irrelevant to cybersecurity. We then manually read most of the articles ($>$70\%) from each of those topics and removed the ones whose contents were  irrelevant. The statistics about the sanitised dataset is demonstrated in Table \ref{t_corpus_analysis}. %More details about the collected articles are explained in Appendix~\ref{a_statistical_detail}. 
We observed that the number of published articles per month in the 2000s is far smaller than the number in the later ten years, as detailed in Appendix \ref{a_no_published_articles}. Therefore, we mainly focused on the analysis of trends in the 2010s in our study.

\subsubsection{Security Categories Identification}
We carefully examined each topic generated by LDA and reviewed the texts, but found the topics were  not still satisfactory %to identify our dataset in terms of distinction and comprehensiveness. 
because they could not cover the dataset completely and had overlapped excessively.
To solve this issue and provide more in-depth insights, we further identified the security categories with term extraction and did manual category identification by card sorting instead of using topics generated by LDA \cite{spencer2009card}.

\textbf{Term extraction}.
We extracted terms from the articles of each topic separately. We employed TermSuite \cite{cram2016terminology}, a toolkit to identify (multi-word) term variants where the termhood is measured by the relative frequency in a domain-specific corpus as well as a general corpus. We only kept the candidate terms  with measure values higher than $2$, a threshold recommended by \cite{cram2016terminology}. %RTo select the representative terms, we removed the ones which occurred in less than 10\% of the input articles. This provided the most reasonable number of terms to represent each topic comprehensively ($Mean: 46$, $SD: 23$).
As a result, we have a collection of terms to replace and represent each topic generated by LDA, with a reasonable number of terms per topic ($Mean: 46$, $SD: 23$).

\begin{table*}[htbp]
\centering
\caption{The 16 manually classified security categories.}
\label{t_categories_identification}
\begin{tabular}{ll|l|l}
\hline
\multicolumn{2}{l}{\textbf{Category}}  & \textbf{Definition} & \textbf{Example terms} \\ \hline\hline
\textbf{CycmnAc} & cybercriminal activity & The malicious activity where the hacker group & malicious action, hacker,\\
&&leverages computer techniques for illegal purposes.  &law enforcement action  \\\hline
\textbf{CysePrg} & cybersecurity program & The cybersecurity venue or event hosted by an authorised organisation, & cybersecurity conference,\\
&&e.g. awareness training, foundations learning, risk assessment. & CISO Forum, consumer education \\ \hline
\textbf{ElecSe} & election security & The protection of elections and voting infrastructure from cyber attack, & voting security,\\
&&e.g. tampering with or infiltration of voting machines and equipment, &election system,\\
&& election office networks and practices, and voter registration databases. & electronic voting machine \\\hline
\textbf{FMClm} & false/misleading & The deceptive advertising claimed by business online, & deceptive claim/advertising, \\ 
& claim& illegal claims about product quality, condition, or price &online complaint assistant\\\hline
\textbf{IdtFncFrd} & identity theft & Criminals gain unauthorised access & data breach, financial crimes,\\
&/financial fraud&to steal credentials to cause unintended charges.  & credit card fraud \\\hline
\textbf{InfPry} & information privacy & Actions that harm or protect users' privacy preferences & customer privacy/data,\\
&&and personally identifiable information. & GDPR, privacy protection \\ \hline
\textbf{IoTThr} & IoT threat & Security threats in IoT devices, software & firmware, mobile device,\\
&& and network connected to the internet. & industrial control systems \\ \hline
\textbf{MalVr} & malware/virus & Malicious software developed to harm computers or networks. & spyware, adware, worm, trojan \\ \hline
\textbf{MbAppSe} & mobile/application  & Security solutions or attacks at the software level, &  fake android app, mobile security, \\
&security& e.g. android apps. &mobile-threat report\\\hline
\textbf{NatSe} & national security & The security and defence of a nation-state, & cyberespionage,\\
&&e.g. its citizens, economy, and institutions, & national cybersecurity,\\ 
&& which is regarded as a duty of government. & transnational crime \\ \hline
\textbf{NetAtk} & network attack & Malicious attempts to gain the unauthorised privilege of network & DDoS attack, zombie bot, \\
&&or cause service disruption. & remote code execution\\\hline
\textbf{PwdEnc} & password/encryption & Password/data protection and encryption.  & MFA, RSA encryption \\\hline
\textbf{SeSwServ} & security software & Software or services designed to help users against attacks & antivirus, MalwareBytes, SIEM, \\
& /service& e.g. antivirus products, educational services & security company,malware filter\\\hline
\textbf{SeUdVnb} & security update & Security weakness exploited by hackers to perform malicious activity. & flaw, patch, security bulletin \\
&/vulnerability & Security update fixes the system or application bugs. &Microsoft Exploitability Index\\\hline
\textbf{SpmPh} & spam/phishing & Scammers spread unsolicited messages online or in social media  & scam, identity parameter, \\
&& with malicious links to steal sensitive information or infect computers. & spam, junk/phishing email \\\hline
\textbf{WbAtk} & web-based attack & Malicious action on web browsers, extensions and content management, & SQL injection, web extension, \\
&& e.g. leveraging third-party plugins to perform code injection. & drive-by download \\\hline
\end{tabular}
\end{table*}

\textbf{Category identification}.
We identified the security categories of our  texts based on the lexical semantics of the generated terms. After removing the duplicates, we applied open card sorting \cite{spencer2009card} to categorise the 810 terms across all the topics. We randomly selected 100 terms and classified them into different categories, and  then applied those to the rest and kept identifying new categories. In total, 16 categories were identified whose details are given in Table \ref{t_categories_identification}. We borrowed the abbreviation styling method of  \cite{rader2015identifying} for the topics.

We further applied the 16 categories to all the terms, where each term  assigned to a maximum of three categories. Three researchers from our faculty who had expertise in cybersecurity performed a manual classification. Each term was classified following the rule of majority voting \cite{narasimhamurthy2005theoretical}. We used Cohen's Kappa \cite{cohen1960coefficient} to measure the agreement between each pair of labellers. The resultant values (all $>$ 0.93) indicate strong agreements between the labellers. An expert review was conducted to ensure the validity of the classification. We recruited two experts in a governmental research lab who had at least three years of experience in the cybersecurity field for this purpose.  For each expert, we generated a 200-term sample (25\%) to review, while our researchers were sitting next to the expert to respond to any questions based on the think-aloud protocol \cite{krombholz2017have}. After our explanations, there was only one error correction, that merely added one term to one more category.

We built a corpus for each category as a set of terms. Each term was added into the corpus of its assigned categories. We used the corpus to measure the relevance of each category to the documents. We identified the duplicate terms semantically or syntactically in the corpus and labelled them as term variants. For instance, `infected computer' is a variant of `infected machine', just as `sensitive data' is a variant of `sensitive information'. Fig.~\ref{f_term_clean} in Appendix \ref{a_no_terms_clean} shows the number of the terms in each category after extending the definition of duplicates to accommodate variants. We see that  security solutions or attacks, especially those related to sensitive information  (\textit{information privacy}, \textit{security software/service} and \textit{security update/vulnerability}) have the largest corpora of terms. In contrast, political or nationwide threats (\textit{election security} and \textit{national security}) contain fewest terms. %as they emerged mainly in the last decade in which hacking processes were described in more general ways with less variety in  techniques. For example, cyber criminals tried to affect election results by penetrating the voter registration system or database.

\subsubsection{Metrics and Analysis}
Instead of using the topic probability computed by LDA, we define the category relevance based on our identified security categories. We obtained a set of term corpora for $K$ categories as $\boldsymbol{C}=\{C_{1}, C_{2},\dotsc,C_{K}\}$.

\textbf{Category relevance}. 
%We define the category relevance of a document as the distribution of unique terms over the categories. 
The category relevance of a document measures the proportion of terms in each category corpus that occur in the document. More specifically, the relevance of a document to each category is computed as,
%We define the relevance of a given document $d_{i}$ to each category $C_{k}$ as
\begin{equation}
\label{e_cate_relev}
\gamma(d_{i},C_{k})=\frac{\abs{c_{k}}}{\abs{C_{k}}}, 1\leq k\leq K%c \in C, c \in d_{i}
%c=(C\cap d_{i})
\end{equation}
%\begin{equation}
%\theta(d_{i})=(\theta(d_{i},C_{1})), 1\leq k\leq K
%\end{equation}
where \abs{C_{k}} is the number of terms in a category corpus, and $c$ denotes the subset of $C$ whose terms occur in the document $d_{i}$. A term is counted once whether it or its variants occur in the document. %The category relevance %of each document is represented as a probability vector with $K$ dimensions. 

\textbf{Dominant categories}.
As explained in \cite{wan2019discussed}, we define the dominant categories of each document as
\begin{equation}
\label{e_dom_cate}
dc(d_{i})=\{C_{k}\}, if \gamma(d_{i},C_{k})>\theta(C_{k}), 1\leq k\leq K%if \max(\theta(d_{i},C))=\theta(d_{i},C_{k}) \forall C \in \boldsymbol{C}
\end{equation}
where $\theta(C_{k})$ is the threshold to determine whether a category is dominant or not. Each document can have different dominant categories. The concept of dominant categories enables us to classify the documents based on the values of relevance.  

\textbf{Category popularity}.
We applied the measures of popularity and impact defined in \cite{wan2019discussed} on the categories. We define the popularity for the category $C_{k}$ within the dataset $D$ as,
\begin{equation}
\label{e_cate_pop}
popularity(D,C_{k})=\frac{\abs{\{d_{i}\}}}{\abs{D}}, d_{i}\in D, C_{k} \in dc(d_{i})
\end{equation}
The popularity of a category measures the proportion of documents with the given category as dominant.

\textbf{Category impact}.
The absolute and relative impact of the category $C_{k}$ is defined as,
\begin{equation}
\label{e_abs_imp}
impact_{absolute}(D(month),C_{k})=\sum_{d_{i}\in D(month)}\gamma(d_{i},C_{k})
\end{equation}
\begin{equation}
\label{e_rel_imp}
impact_{relative}(D(month),C_{k})=\frac{impact(D(month),C_{k})}{\abs{D(month)}}
\end{equation}
where $D(ts)$ represents a collection of documents posted in a month. The absolute impact of a category measures the cumulative relevance to the category of the posted documents over a month. The absolute impact is influenced by the number of posts and their category relevance. The relative impact is not affected by the number of posts. It measures the average relevance to the category of the posted documents during a month.

%surya - the missing piece here is the intuition behind each metrics. What do they measure? What do it means when we say impact is high? etc. We need to justify why these metrics are chosen and the reasoning behind the choice. 
%tingmin - explained what do they measures and the intuition

\section{Results}
\label{s_result}
We exhibit our results following our methodology in this section. Through data analysis, we try  to answer our three research questions.

\subsection{RQ1. What are the security issues reported in cybersecurity texts?}

We first generated 68 topics out of our datasets by using LDA; 18 from news, 15 from security blogs, and 35 from the three web datasets, as shown in Table \ref{t_corpus_analysis}. However, the LDA-generated topics were not good representatives and were hard to distinguish too. Therefore, we further studied the terms from the topics and manually found 16 security categories that can represent the articles, as demonstrated in Table \ref{t_categories_identification}. The table explains each security category with examples in detail. The categories were identified based on different perspectives on cybersecurity, including attack types (e.g. network, web, IoT, or mobile/application attacks), security techniques (e.g. encryption and security services as well as updates) and recently emerged security issues (e.g. election and national security).
%surya - you need to say what insights you gained from 16 categories. I think the results should be more than a report of the numbers. What do they mean? What they represent. Three studies below so the validation of the results, but it lacks a punch line. For example, it demonstrates that privacy is a dominant topic discussed in the last two decade. Password has been an issue - no wonder usable authentication is still remains a challenge. We need to give a concrete message from the results. The results provided in the textbox are quantitative results from the experiments. I think we need to give a meaning to those results. Otherwise, we are reporting the results with not much insights. It would be good if you can provide a paragraph for each RQ, presenting the interesting findings related to attacks, challenges - that inspires researchers to do further research or provide hot topics. 
%tingmin - 
% 1. added the insights in Category co-occurrences

\textbf{Category validation}. We validated the effectiveness of our 16 security categories. We applied the chi-squared (${\chi}^2$) test on the consistency of  each category prevalence. More specifically, we tested if the proportions of the articles were similar with a given category as the most relevant category. %first, second or third topic. We defined the dominant 
The most relevant category of a document is the category where it achieves the highest relevance (Equation \ref{e_cate_relev}). %The second and third dominant topics are the ones with the next two largest $\gamma$s. 
Table \ref{t_p_security_categories} in Appendix \ref{a_test_category_validation} shows the results of ${\chi}^2$ tests for each category across 67 LDA-generated topics in the five datasets from the three sources. As all the $p$ values are smaller than 0.01, the prevalence of our classified categories is significantly varying across the LDA-generated topics, datasets and sources. It indicates our identified categories are representative and effective because the differences between them are statistically significant. %We further tested the proportions of the articles with a given category as its first dominant topic only, and results are similar (all $p<0.01$).

%surya - in the category co-occurrence, we need to explain why are we doing this. What was the aim? What it means to researchers and practitioners. I am happy with the results presented, but it lacks the revelation.  
%tingmin - added the aims and insights below.

%\begin{figure}[htbp]
\begin{figure}[t]
\centering
\includegraphics[width=1\linewidth]{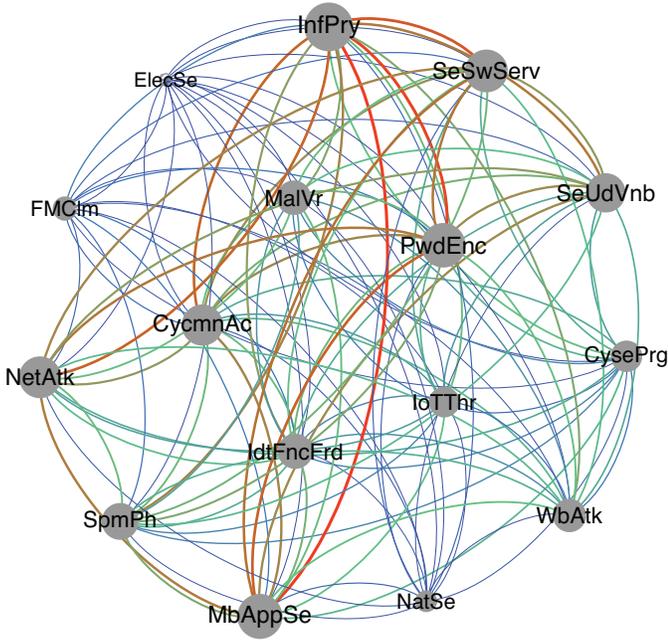}
\caption{The network graph of co-occurrences between different categories within an article. The size of the nodes represents the relative occurrences of each category. Red (thick) lines mean strong relationships (high co-occurrences) between two nodes, while blue (thin) represent weak relationships.}
\label{f_network_co_occurrence}
\end{figure}

\textbf{Category co-occurrences}. We explored the relationships between different categories by calculating their co-occurrences in each document. The co-occurrences show the associations between different security issues to pinpoint the challenges faced by researchers and practitioners. Fig.~\ref{f_network_co_occurrence} presents the network graph of the co-occurrences, where larger nodes indicate more frequently occurred categories and red (thicker) lines show strong relationships. 
We find that \textit{information privacy}, \textit{password/encryption}, \textit{mobile/application security} and \textit{network attack} are strongly correlated. These four categories have also had high numbers of occurrences in our dataset. \textit{cybercriminal activity} exhibits strong relationship with \textit{information privacy}. This indicates that information privacy still remains a dominant topic in the last decade and is largely due to criminal offence, including password attack (e.g. brute force attack), mobile application attack (e.g. malicious code injection exposure), and network attack (e.g. DDoS attack). Yet, usable authentication methods, mobile security solutions, and network protection are still challenges in safeguarding the sensitive data (e.g. credentials) of users and enterprises. In addition, the strong correlations between \textit{spam/phishing} and both \textit{network attack} and \textit{mobile/application security} denote that spam and phishing messages including malicious links are still spreading rampantly in the internet through email, SMS or other communications.

Among the 16 categories, \textit{election security} and \textit{national security} occur the least frequently and have the weakest correlation with the rest of the categories. The articles in these two categories mainly present nationwide attacks and espionage at high levels, with a focus on the infrastructure and attack consequences. They hardly analyse the related techniques in detail. Since the targets of these threats, such as governments, are harder to compromise compared to regular users, the attacks are not frequent. However, they are to be taken seriously since they can cause significant losses such as political or military information leakage.
Compared to these two categories, \textit{false/misleading claim}, \textit{IoT threat}, \textit{web attack}, \textit{cybersecurity program} happen more regularly but have weaker connections to other categories. Specific attacks  such as  \textit{web attack} and \textit{IoT threat} are partly related to a few categories. For instance, criminals can leverage cross-site scripting (XSS) (web) attacks to inject malicious codes into web applications (\textit{mobile/application security}) and access sensitive information (\textit{information privacy}).

%\begin{figure}[htbp]
%\centering
%\includegraphics[width=1\linewidth]{proportion_3dt.eps}
%\caption{The proportions of articles in the whole dataset with each security category as their first, second and third dominant topics.}
%\label{f_proportion_3dt}
%\end{figure}

\begin{figure}[t]
\centering
\includegraphics[width=1\linewidth]{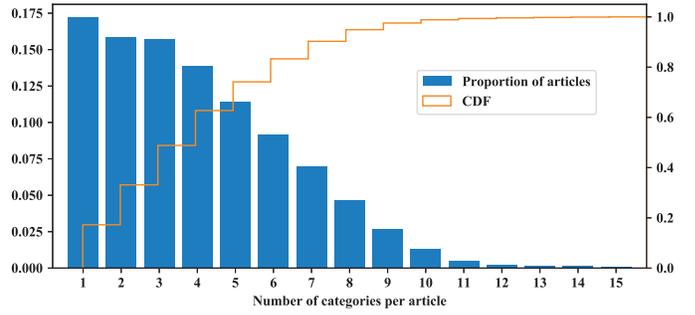}
\caption{Proportion of articles with different dominant categories and the CDF (Cumulative Distribution Function) vs the number of dominant categories (per article).}
\label{f_no_categories}
\end{figure}

\textbf{Categories per article}. We empirically found the threshold per category (Equation \ref{e_dom_cate}) to determine the dominant categories for each article in our dataset. In Fig.~\ref{f_no_categories}, we have plotted the probability distribution of the number of dominant categories for the  articles along with its  CDF (Cumulative Distribution Function). With the increase in the number of dominant categories, the number of articles gradually decreases. The results show that 83\% of the articles have six dominant categories or less. This percentage reaches 90\% with seven categories. 
% surya = what does this mean? We fail to explain the insights here. 
%tingmin - the insights are explained in the following paragraph. I also added a sentence at the end of the paragraph.
The results are aligned with our observations. In practice, different from other fields, these articles generally include multiple topics. For example, when an article introduces cyber attacks and prevention methods, it always explains the techniques and the related effects in detail. For example, a security update addresses an exploitable vulnerability through which remote code execution by hackers is possible. Hackers use this to gain admin access and run malware on infected computers. In this scenario, \textit{network attack}, \textit{password/encryption}, \textit{malware/virus} and \textit{security update/vulnerability} are discussed. %As shown in Fig.~\ref{f_no_categories}, 63\% of articles have up to four dominant categories.
It also indicates that a security article generally discuss multiple security issues.
%\newline
\begin{tcolorbox}[
  %enhanced,clip upper,%<------------
  colframe=black,colback=white,boxrule=1pt,arc=0pt,
  boxsep=1pt,left=1pt,right=1pt,top=1pt,bottom=1pt]
\begin{itemize}[leftmargin=*]
\item We identified 16  categories for the security articles from news,  blogs and webs.
%\item \textit{information privacy}, \textit{password /encryption}, \textit{mobile /application security} and \textit{network attack} occur frequently in our dataset and are strongly correlated;
\item Information privacy still remains a dominant topic in the last decade and is largely due to criminal offence, including password attack, mobile application attack, and network attack. %Yet, usable authentication methods, mobile security solutions, and network protection are still challenges in safeguarding the sensitive data of users and enterprises.
\item Most of the articles (83\%) have six dominant categories or less. 
\end{itemize}
\end{tcolorbox}

%\textbf{RQ2.} How do the security topics evolve in the past 15 years?

\subsection{RQ2. How have the security categories varied and evolved over the last decade?}

\subsubsection{Category Popularity}
We compared the popularity of different categories. We empirically found the threshold to separate dominant categories (Equation \ref{e_dom_cate}) and calculated the category popularity too (Equation \ref{e_cate_pop}). Fig.~\ref{f_popularity_q2} in Appendix \ref{a_category_popularity} plots the popularity of the security categories. \textit{cybercriminal activity} marks the most substantial category amongst all (with 65\% popularity). This category contains the terms indicating cyber attacks such as `hack'. The three categories  \textit{information privacy}, \textit{security software/service}, and \textit{cybersecurity program} share  similar popularities, at around 40\%. In contrast, other categories are discussed less popularly, such as articles introducing specific threats (e.g. spam/phishing, malware/virus). \textit{election security} stands lowest in terms of popularity among the 16 security categories.

%surya - I fix the typo directly in the above paragrpah. 
%tingmin - checked.

\subsubsection{Category Absolute Impact}
%\subsubsection{Impact Comparison between categories}

\begin{figure*}[htbp]
\centering
\includegraphics[width=0.9\linewidth]{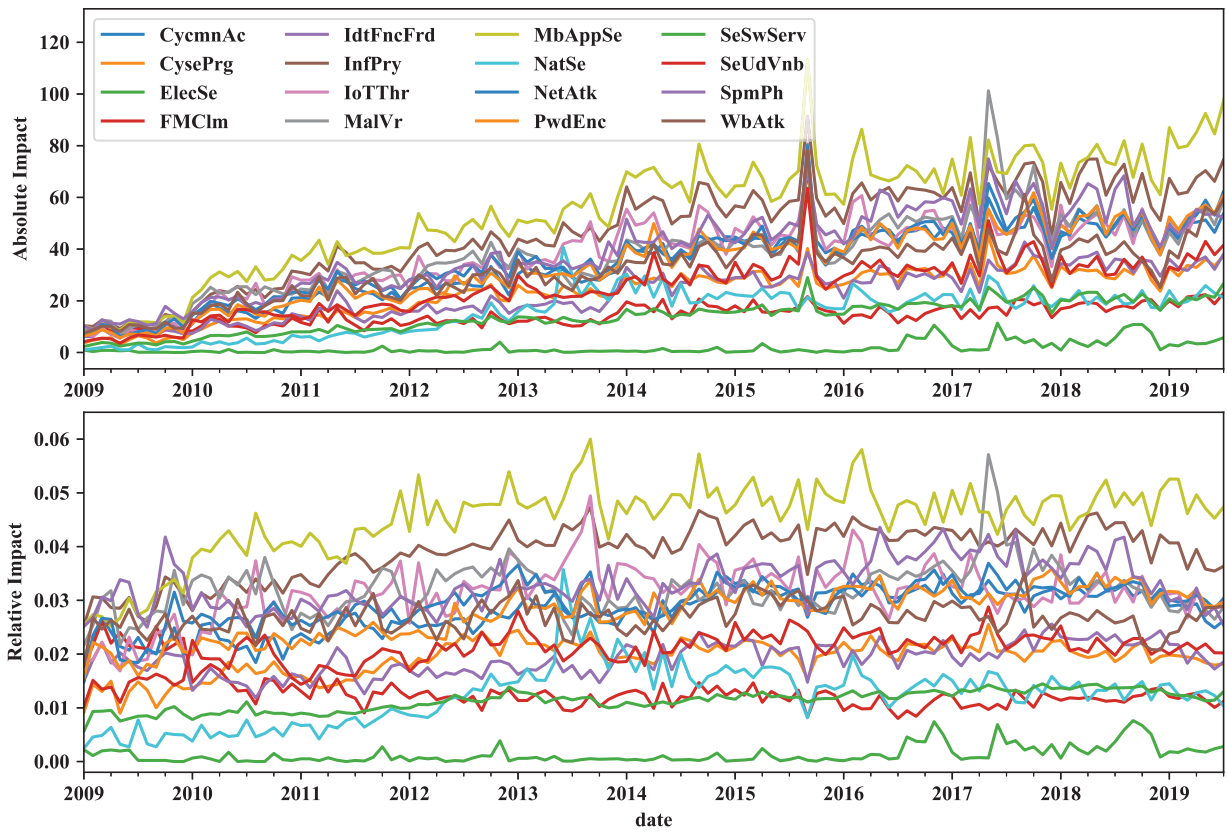}
\caption{The absolute impact and relative impact of 16 security categories from 2009 to date.}
\label{f_impact_q2}
\end{figure*}

\begin{figure*}[htbp]
\centering
\includegraphics[width=1\linewidth]{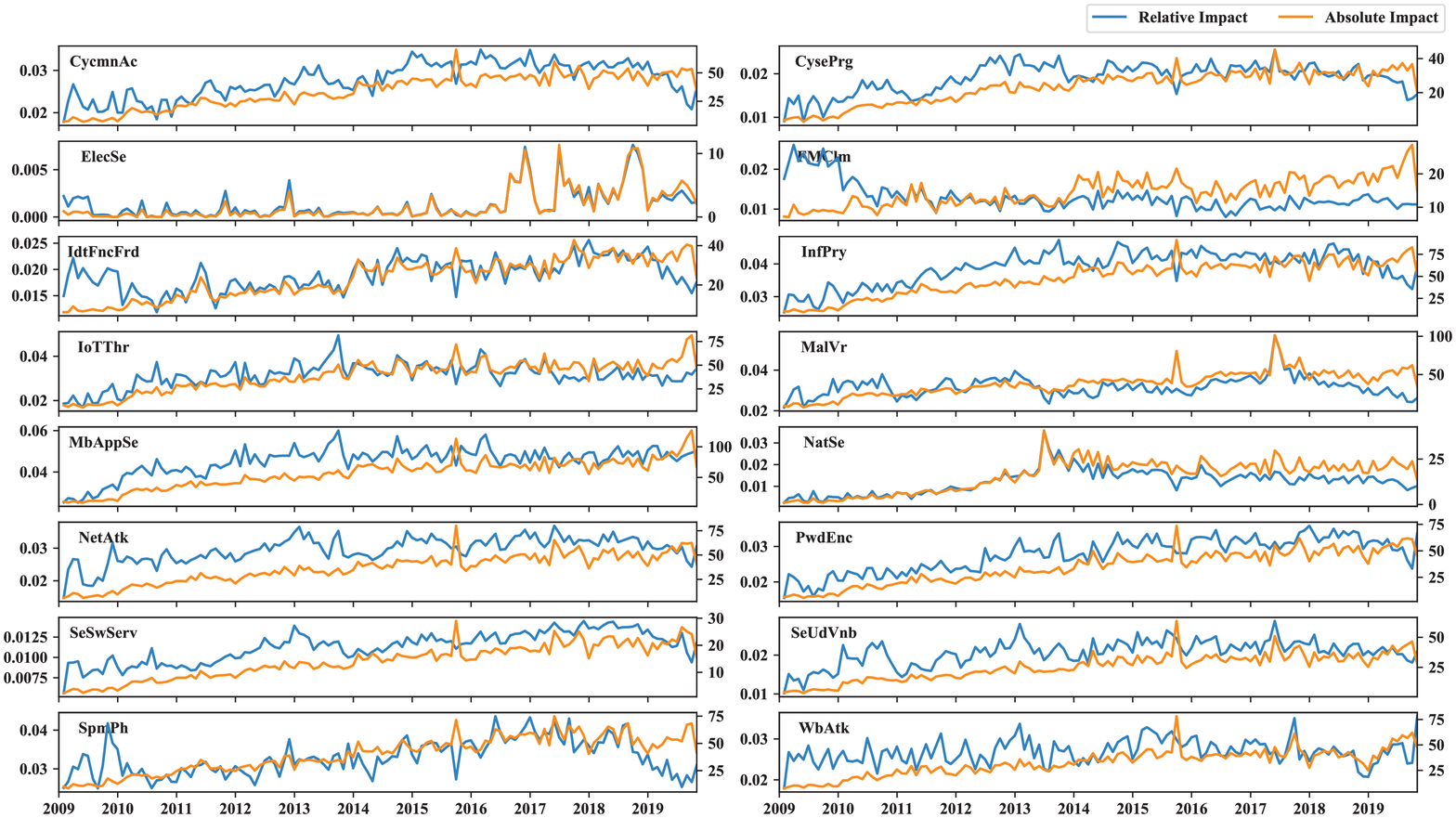}
\caption{The separate absolute impact and relative impact of 16 security categories over the last decade.}
\label{f_impact_q2_split}
\end{figure*}

We calculated each category absolute impact (Equation \ref{e_abs_imp})  to analyse the trends. The results of the last decade analysis are plotted in the upper subplot of Fig.~\ref{f_impact_q2}. Fig.~\ref{f_impact_q2_split} demonstrates the comparison between the trends of absolute impact and relative impact per category. Overall, there is an upward trend in the absolute impact for almost all categories since 2009. The impacts of all  categories start from nearly zero in 2009. The increase indicates a considerable evolution of security incidences in both amount and sophistication. The explosion of ransomware in 2017 brings the  impact of \textit{malware/virus} to its peak, especially with the worldwide break out of WannaCry which infected 200,000 computers across 150 countries \cite{wannacry_effect}.  %The trend ends with a dramatic decrease in the last month, in the middle of which we finished data collection. 

%--------- SAYAD : I RESUMED FROM HERE ---------------

We aggregated the overall absolute impact for all the categories and compared it to the monetary damage caused by cybercrimes in the 2010s (data from \cite{cybercrimedamage}). We used Spearman correlation coefficient to measure the correlation between the overall absolute impact of security articles and the amount of financial loss caused by recorded cybercrimes. The result showed a strong correlation ($corr=0.85, p=0.0037$). The increasing impact of %evolved cyber incidents
security categories reflects exponential economic loss, from \$0.3 million in 2015 to \$3.5 billion in 2019.

%Surya - the paragrph below is a good one. I would like to have a similar one in RQ1; we need to provide the meaningful observations. 
We observe that there was a sharp jump in the absolute impact at the end of 2015 for most categories, followed by another steady growth in 2017. Interestingly, almost all the categories had  decreasing trends in absolute impact in 2018, but climbed to the highest point in 2019. Different from other categories, \textit{election security} had a significant increase in the absolute impact in 2016, while it was around zero before that. This increase coincides with the Russian interference in 2016 U.S. presidential election \cite{2016_US_election}. \textit{national cybersecurity} became popular earlier, with the absolute impact gradually going up from 2009, before a sudden rise in 2013. National security, including national cyber attacks and cyber-espionage, was first considered to be more harmful than other threats (e.g. terrorism) by U.S. officials in 2013 \cite{2013_national_security}. It is worth noting that absolute impact and relative impact almost overlap for both \textit{election security} and \textit{national security}, as shown in Fig.~\ref{f_impact_q2_split}.

\subsubsection{Category Relative Impact}

We additionally computed  category relative impacts (Equation \ref{e_rel_imp}) for the sake of comparison. Relative impact reflects the average impact that each article has on the security categories during a month. The results of studying the articles from 2009 to now are depicted in the lower subplot of Fig.~\ref{f_impact_q2} and Fig.~\ref{f_impact_q2_split}. %The separate relative impact is depicted in 16 subplots of Fig.~\ref{f_impact_q2_split}. 
Among the 16 categories, \textit{mobile/application security} and \textit{information privacy} have had the largest relative impacts over time as well as the largest absolute impacts. Meanwhile, \textit{election security} shows the smallest relative impact and absolute impact. 

We computed the Pearson correlation coefficient of all the pairs relative impacts from the security categories. We found the trends of the relative impacts for seven categories (i.e. \textit{cybercriminal activity}, \textit{cybersecurity program}, \textit{information privacy}, \textit{network attack}, \textit{password/encryption}, \textit{security software/service} and \textit{spam/phishing}) are similar to each other ($corr>0.7, p<0.01$). The relative impacts of them have progressively risen from 2009 to 2015, and fluctuated around the peak afterwards.

We further applied the Mann-Kendall trend test \cite{Hussain2019pyMannKendall} to statistically measure the trends of relative impacts for the categories. The results suggest that 15 out of our 16 categories have   statistically increasing trends ($p<0.05$). Only one category (\textit{false/misleading claim}) experiences a  downtrend  ($p=1.87e-11$). 

%Fig.~\ref{f_impact_q2_split} demonstrates the comparison between the trends of absolute impact and relative impact per category. We observe that  absolute impact and relative impact almost overlap for both \textit{election security} and \textit{national security}. The impact of \textit{election security} shows a significant increase after 2016, while before that the impact was nearly zero. Similarly, the impact of \textit{national security} has steadily  grown since 2009 and sharply jumped to the peak in 2013. The impact declined slightly in the same year but remained steady in the next six years. 
%Surya - in the above paragrph, it would be good if we can relate to the US election; it would be good to provide the links between the results and real life events. You can probably dig in and look at the articles and relate to the real events. For example, US election results in higher election or national security article. After the target attacks, there is a spike in information security or crime. You need to look at the articles and relate to the real events to make the study more meaningful to the readers/reviewers.
% tingmin - election security and national security have been explained at the end of Category Absolute Impact. So I deleted the above paragraph.

%\newline
\begin{tcolorbox}[
  %enhanced,clip upper,%<------------
  colframe=black,colback=white,boxrule=1pt,arc=0pt,
  boxsep=1pt,left=1pt,right=1pt,top=1pt,bottom=1pt]
\begin{itemize}[leftmargin=*]
%\item \textit{cybercriminal activity} has the largest popularity (65\%) among the 16 categories, followed by \textit{information privacy}, \textit{security software/service} and \textit{cybersecurtiy program}, with similar popularities at 40\%;
\item \textit{cybercriminal activity} has been the most popular and was discussed in most security articles (65\%), followed by \textit{information privacy}, \textit{security software/service} and \textit{cybersecurtiy program}, with similar popularities at 40\%.
%\item \textit{mobile/application security} and \textit{information privacy} had the largest absolute impact as well as relative impact over the last decade; 
\item Almost all the categories show upward trends in both absolute impact and relative impact over the last decade.
\item Security issues in mobile/application and information  privacy gained the largest absolute/relative impact over time.
\item The absolute impacts %extracted 
from cybersecurity texts strongly correlate with the monetary loss caused by cybercrimes.

\end{itemize}
\end{tcolorbox}

\subsection{RQ3. How have the security categories varied and evolved across different sources on cybersecurity over the last decade?}

\begin{figure*}[htbp]
\centering
\includegraphics[width=0.6\linewidth]{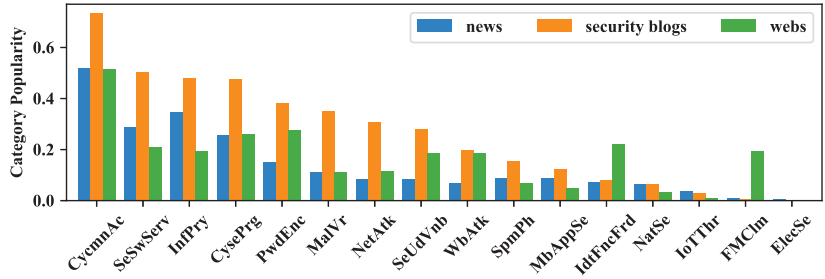}
\caption{The popularity of our security categories in different sources about cybersecurity.}
\label{f_popularity_q3}
\end{figure*}

\begin{figure*}[htbp]
\centering
\includegraphics[width=1\linewidth]{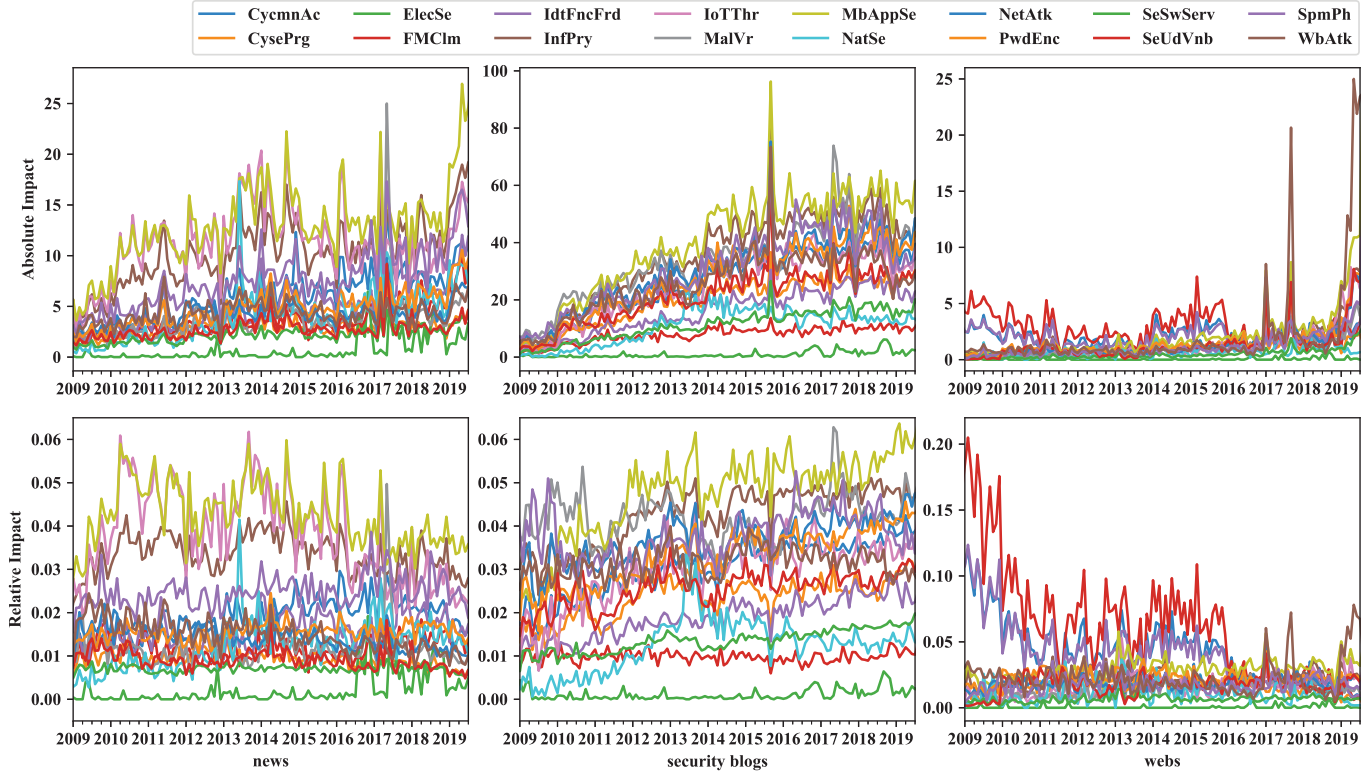}
\caption{Absolute and relative impacts of our security categories for different sources of cybersecurity from 2010 to date.}
\label{f_impact_q3}
\end{figure*}

\begin{figure*}[htbp]
\centering
\includegraphics[width=1\linewidth]{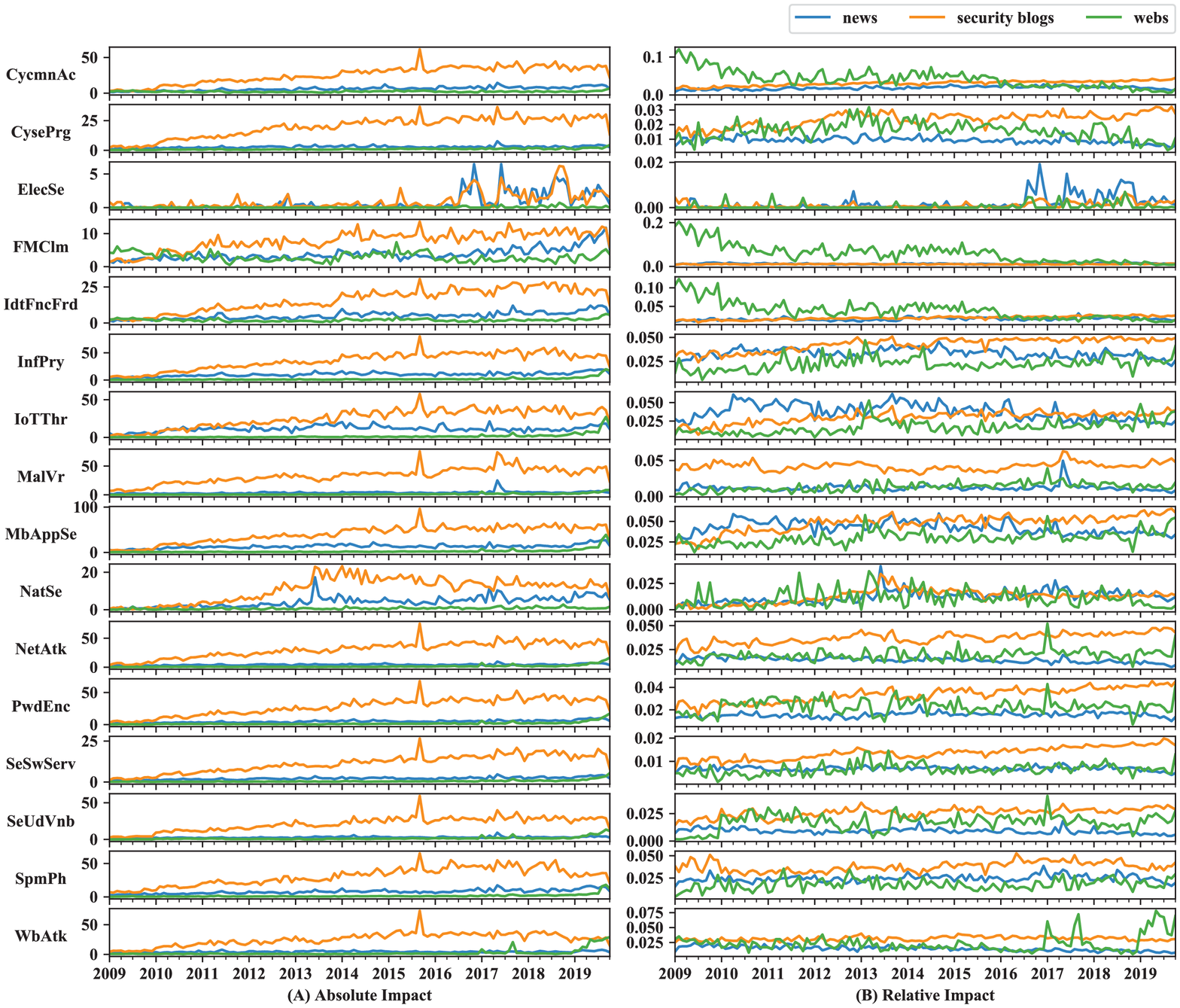}
\caption{The (a) absolute impacts and (b) relative impacts of our 16 security categories across various sources (news, security blogs, webs) over the last decade.}
\label{f_impact_q3_impact_split}
\end{figure*}

We compared the security categories in terms of their popularity and impact across different sources of cybersecurity articles, i.e. news, security blogs and websites. This provides insights on how categories become popular  on different  platforms. %Comparison of impacts reveals different trends of security impacts across different sources.

\subsubsection{Category Popularity}
Fig.~\ref{f_popularity_q3} demonstrates the category popularity of security articles across the three sources. We find that almost all the categories are popularly present within all the sources except \textit{election security} and \textit{false/misleading claim}. \textit{election security} only presents its prevalence in news at a significantly lower popularity compared to other categories.  \textit{false/misleading claim} refers to fake or deceptive online advertisements designed to mislead customers. This category is  mainly active in web pages, but has shallow popularity in news.
Among the 16 categories, \textit{cybercriminal activity} has the highest popularity in all the sources. It is worth noting that among the three sources, security blogs stand popular in the majority of  categories. This is because security blogs are more domain-specific and contain more detailed security knowledge in the content. Interestingly, only four categories (i.e. \textit{identity theft/financial fraud}, \textit{password/encryption}, \textit{security update/vulnerability}, and \textit{website attack}) show considerably higher popularity in web sites than in news.

\subsubsection{Category Absolute Impact}
The absolute impact for the security categories at different sources are depicted in the upper plot of Fig.~\ref{f_impact_q3} and Fig.~\ref{f_impact_q3_impact_split}(A). Overall, there is an increasing trend in the absolute impacts for all the three sources. We further used the Mann-Kendall trend test \cite{Hussain2019pyMannKendall} to check whether the trend is statistically significant or not. The results show that the increasing trends of absolute impacts for all the categories in news and security blogs are significant ($p<0.05$). In webs, two categories (i.e. \textit{cybercriminal activity} and \textit{identity theft/financial fraud}) do not have any significant trends ($p=0.06, 0.2$), whether increasing or decreasing. Only \textit{false/misleading claim} experiences a significant downtrend ($p=0.008$) in absolute impact during the 2010s.

From Fig.~\ref{f_impact_q3}, we observe that the distinction in the absolute impacts of different categories is less significant in webs than in the other two sources over time. Overall, security blogs have the largest absolute impact among the three sources. The high value of  absolute impact indicates that security blogs have been the dominant source of delivering security knowledge in the last ten years. Compared to the other sources, web sources have had low absolute impacts ever since 2009, however, the trend has ended with a dramatic increase in 2019.
Among the security categories, \textit{mobile/application security} has gained the highest absolute impact, especially in news and security blogs. Besides, \textit{information privacy} has achieved the second highest absolute impact across all the sources at almost  all times. In news, its absolute impact exceeded \textit{IoT threat} and moved up to the second after 2016. In webs, it surpassed \textit{mobile/application security} and reached the highest in the first half of 2019.

Fig.~\ref{f_impact_q3_impact_split}(A) plots the value of absolute impact  for each category separately. In security blogs, we observe that most categories experience a rapid rise in the absolute impact in 2015. Except for \textit{election security} and \textit{national security}, the trends of the remaining categories are similar; with a steady increase at different paces. Compared to security blogs, news and web pages gain considerably lower absolute impacts, except for \textit{election security}. Moreover, news absolute impact is slightly higher than that of web pages.

\subsubsection{Category Relative Impact}
The lower plot of Fig.~\ref{f_impact_q3} and Fig.~\ref{f_impact_q3_impact_split}(B) demonstrate the relative impacts of the security categories across the three sources. They show that   \textit{mobile/application security} has had the largest relative impact in news and security blogs during the study period. This is similar to its absolute impact. It is worth noting that \textit{IoT threat} almost mirrored the absolute impact and the relative impact of \textit{mobile/application security} in news. Interestingly, \textit{malware/virus} rarely made a higher relative impact than \textit{mobile/application security} in security blogs. One can also see that  \textit{false/misleading claim}, \textit{identity theft/financial fraud} and \textit{cybercriminal activity} have had the highest relative impacts in webs before 2016. 

From Fig.~\ref{f_impact_q3_impact_split}(B), we find that the relative impact of web pages fluctuates  dramatically. While security blogs still took the dominant place in relative impact for most the categories during the study period, it was taken over by news and web pages sometimes. Webs maintained the highest relative impact in \textit{cybercriminal activity}, \textit{false/misleading claim}, and \textit{identity theft/financial fraud} until 2016. Moreover, news showed to have a  relative impact comparable to security blogs in some categories, namely \textit{information privacy}, \textit{national security} and \textit{spam/phishing}. This source took over security blogs in \textit{election security}, \textit{IoT threat} and \textit{mobile/application security} occasionally.
In our collected data, the proportion of web articles with publication dates is significantly smaller than that of news or security blogs, with  percentages around 66.6\% compared to at least 98\% for the latter two. This leads to low absolute impacts in webs, in contrast to  noticeably higher relative impacts compared to the other two sources.

We further applied the Mann-Kendall trend test \cite{Hussain2019pyMannKendall} to see whether the trend of relative impact has been statistically significant over time or not. The results are reported in Table \ref{t_p_rel_impact_q3_trend} in Appendix \ref{a_sig_test_sources}. Only security blogs showed increasing trends ($p<0.05$) in the relative impact for nearly all the categories, except  \textit{false/misleading claim} which was the only category that remained stable. This category showed the same behaviour in the other two sources too. News and webs had a few decreasing trends among their categories. The results also suggest that half of the security categories  experienced  statistically-significant increasing trends ($p<0.01$) in relative impact %at all times
across the three sources %These categories were
(i.e. \textit{election security}, \textit{information privacy}, \textit{IoT threat}, \textit{malware/virus}, \textit{mobile/application security}, \textit{network attack}, \textit{security software/service} and \textit{spam/phishing}).
%\newline
%%%-\begin{tcolorbox}[
  %enhanced,clip upper,%<------------
%%%-  colframe=black,colback=white,boxrule=1pt,arc=0pt,
%%%-  boxsep=1pt,left=1pt,right=1pt,top=1pt,bottom=1pt]
%%%-\begin{itemize}[leftmargin=*]
%\item \textit{cybercriminal activity} has had the largest popularity among the 16 categories across all the sources; 
%\item For the majority of categories, security blogs have been the most popular and had the largest absolute/relative impacts among the sources over the last decade;
%%%-\item For most categories, security blogs have been the most popular and impactful among the sources in the 2010s;
%\item \textit{mobile/application security} has gained the largest absolute impact as well as relative impact in news and security blogs during the past ten years;
%%%-\item security issues in mobile/application have been the most impactful in news and security blogs over time;
%%%-\item \textit{IoT threat} almost mirrored the absolute  impact value of \textit{mobile/application security} in news over time;
%%%-\item Only security blogs  experienced statistically increasing trends in  relative impact for nearly all the categories.
%%%-\end{itemize}
%%%-\end{tcolorbox}
\begin{tcolorbox}[
  %enhanced,clip upper,%<------------
  colframe=black,colback=white,boxrule=1pt,arc=0pt,
  boxsep=1pt,left=1pt,right=1pt,top=1pt,bottom=1pt]
\begin{itemize}[leftmargin=*]
\item For most categories, security blogs have been the most popular and impactful among the sources in the 2010s.
\item Security issues in mobile/application have been the most impactful in news and security blogs over time.
\item \textit{IoT threat} almost mirrored the absolute  impact value of \textit{mobile/application security} in news over time.
\item Only security blogs  experienced statistically increasing trends in  relative impact for nearly all the categories.
\end{itemize}
\end{tcolorbox}

\begin{figure}[htbp]
\centering
\includegraphics[width=1\linewidth]{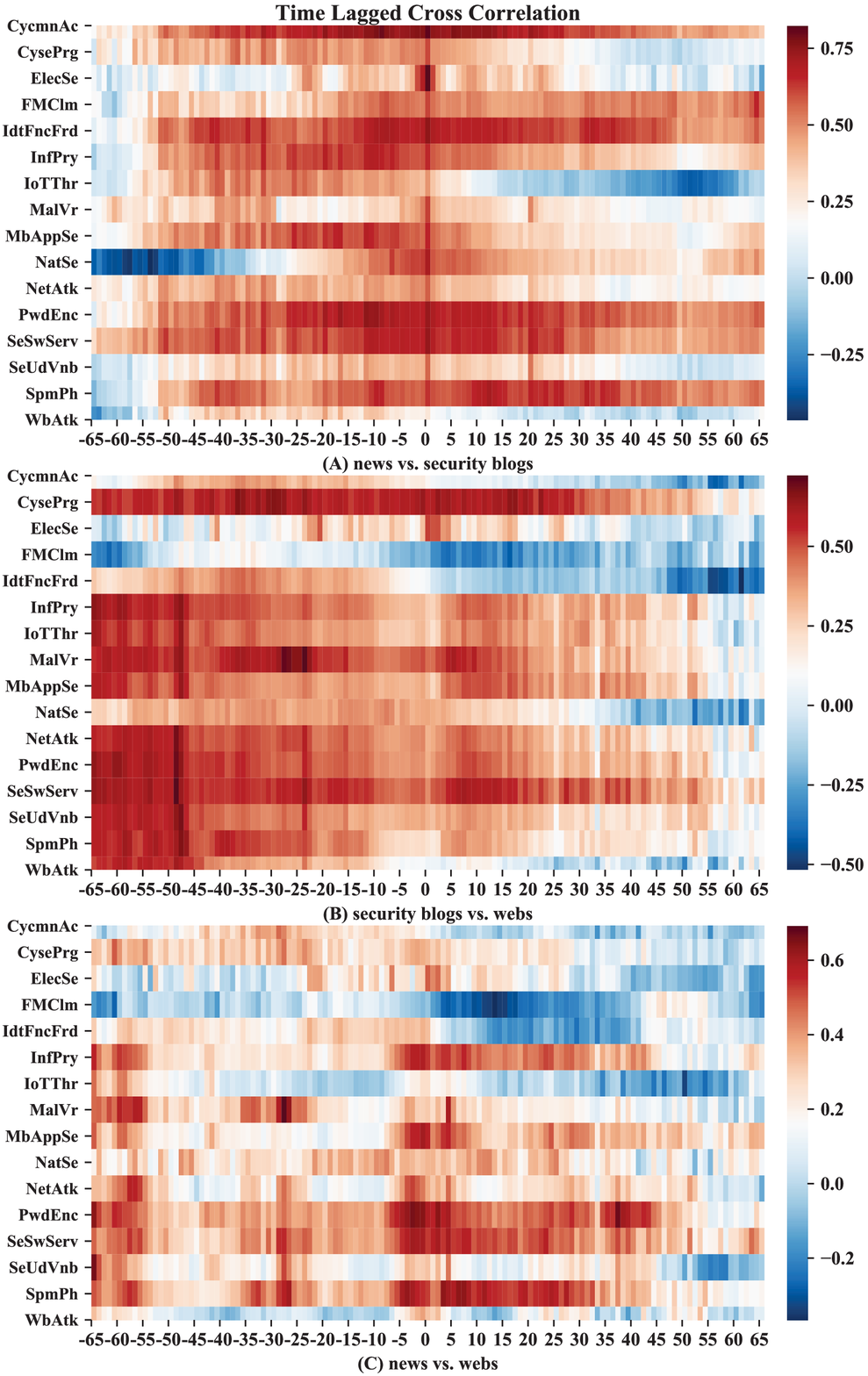}
\caption{The time lagged cross correlations (TLCC)\cite{boker2002windowed} in the absolute impact between (A) news vs. security blogs, (B) security blogs vs. webs and (C) news vs. webs. It plots the correlation between each pair of sources with shifting the second source (after `vs.') backwards ($-$) or forwards ($+$) in months. Darker red colour represents positively stronger correlation, and the peak correlation (dark red) indicates the two sources are most synchronised at that time.}
\label{f_information_flow}
\end{figure}

\textbf{Timeliness of different sources}. 
We further measured the timeliness of different sources in reporting security incidents. We compared the sources temporally to measure the difference in information delivery delay. We applied time-lagged cross correlations (TLCC) \cite{boker2002windowed} to calculate the correlation progressively with shifting one time series incrementally. TLCC identifies any temporal (leader-follower) relationship between two time series. The comparisons between each pair existed in the sources in the absolute impact per category are plotted in Fig.~\ref{f_information_flow}. Each subplot depicts the dynamic correlation when we pull  the second source backward (negative: $-$) or forward (positive: $+$). Darker colours indicate  stronger correlation, with red representing positives and blue representing negatives in the spectrum. The peak correlation (dark red) shows where the two sources are most synchronised in time, either when the first source leads ($-$) or when the second source leads ($+$). 

From Fig.~\ref{f_information_flow}(B), we clearly observe that  'security blogs' drove  'webs' in the study period. A strong correlations is seen if one moves webs backwards for at least 10 months. Web pages show a few month delay compared to news in security information delivery, as shown in Fig.~\ref{f_information_flow}(C). Overall, webs show very weak correlation with news, in contrast to the other two pairs. This indicates that websites do not focus on the timeliness when publishing cybersecurity texts, which is aligned with their low proportion of articles having publication dates. Fig.~\ref{f_information_flow}(A) suggests that news and security blogs report security events firsthand and at almost similar speeds in most categories. News  led in \textit{information privacy}, \textit{IoT threat}, \textit{mobile/application security}, \textit{password/encryption}. Only \textit{spam/phishing} was an exception and became influential  in security blogs earlier.
%\newline
\begin{tcolorbox}[
  %enhanced,clip upper,%<------------
  colframe=black,colback=white,boxrule=1pt,arc=0pt,
  boxsep=1pt,left=1pt,right=1pt,top=1pt,bottom=1pt]
\begin{itemize}[leftmargin=*]
\item News and security blogs report security events firsthand at similar speeds in most categories.
\item Websites deliver security information without caring about timeliness much, where one third of the articles do not specify the date and the rest have a time lag in posting emerging security issues.
\end{itemize}
\end{tcolorbox}

\section{Discussion}
\label{s_discussion}
%\textbf{One-fits-all security education is impractical}. 
\textbf{Security education}.
Home computer users are struggling to resist the ever increasing cyber threats. While  formal security education designed by certified experts is essential, it is still challenging to standardise the training as users might need security knowledge at different levels. Moreover, different users may have different backgrounds in dealing with cyber attacks. %, including  education and work experience. 
That is why informal online sources have become a major platform for users to  learn security advice from. Understanding  cybersecurity texts and the difference between  sources can help users with the identification of useful information by themselves. Our analysis can additionally help to improve  current security information sharing systems by capturing the trending topics with time. In this paper, we only studied three sources (news, security blogs and websites). There are more online information sources in the real world, such as technical reports which are not covered here. However, the three sources we picked are the best representatives in terms of  prevalence, authority and users' click rate. They also broadly cover the security information reported by other sources. %To avoid  redundancies, we only kept these three as the primary datasets for our  research purpose.

%\textbf{Digital source selection to obtain useful information about cybersecurity}.
\textbf{Security advice}.
Apart from online sources, there exist other ways for users to get advice and make security decisions. IT workers, especially those with qualified internet skills who process  sensitive business data are likely to learn from negative experiences too \cite{redmiles2016learned}. Home computer users also gain security knowledge from social learning, such as their family, friends and acquainted experts \cite{das2014increasing,das2014effect,rader2012stories}. Regardless of the diversity in security learning methods, it is hard to collect the real-world information received  from communications and convert it into a standard text format. Thus, in our study, we only considered the online articles with text content that could be easily accessed by end users. Our collected articles from security blogs and webs somehow contained social communications too. Note that security experts are likely to share  security stories and tips to safeguard both home computer users and businesses. People with negative experiences might also post their personal stories in discussion web forums to seek help from authorities as well as other online users. 

\textbf{Cyber attack prediction}.
%We analysed the trending topics from cybersecurity texts. They reveal how  security incidents evolve. Understating the latest reported cyber attacks and the correlation between the security categories involved can help us improve current the cyber attack prediction mechanism.
Criminals are leveraging advanced technologies to perform sophisticated hackings such as %the development of
cryptojacking (cryptomining attacks) based on rapidly grown cryptocurrencies (e.g. blockchain). By studying the topic patterns and following the  tendencies  in existing security incidents, we can predict the security categories that might be exploited by hackers and additionally, infer the potential  technologies to be used. 
Since cybersecurity texts  discuss security issues at different levels of technicality, it is unlikely that one can create a globally-accepted standard set of security topics. %For example, company websites can produce readable and straightforward instructions for  end users such as ``Changing passwords more frequently'', however,  blogs technically explain how hackers steal passwords in detail and make recommendations on the security measures users should employ (e.g. 2FA or two factor authentication). 
Traditional classifications are either too abstract (architecture-based classification, e.g. application layer, endpoint layer) or too specific (common cyber attack types). In contrast to these, our manual classification which that uses card sorting \cite{spencer2009card}, provides a comprehensive set of security categories that cover considerably different levels of security issues. Each article is associated with some categories, which is in line with the observation of real-world data and makes comparison of  articles  discussing even similar attacks, possible. %Further research can explore more robust classification methods for cybersecurity texts.

\section{Conclusion}
\label{s_conclusion}
In conclusion, we discovered the emerging topics in cybersecurity and preformed an empirical analysis based on our collected security texts from the sources of news, security blogs and websites. Since LDA cannot generate specific and distinguished topics for cybersecurity texts, we proposed a novel semi-automated classification method for this purpose. % that leverages the advantages of LDA. 
We applied the term extraction method based on the results generated by LDA. We then identified 16 security categories from the terms that could represent the articles statistically as a probabilistic distribution. We further analysed the evolution and variation of the collected articles across the security categories as well as sources over the last decade. We revealed several interesting findings, like the absolute impact of cybersecurity texts shows a strong correlation with the financial loss caused by cybercrimes, or websites (of authorised organisations), in contrast to news and security blogs, tend to publish general security articles without caring about their timeliness. Further research can be conducted on this subject, such as discovering more comprehensive topics for cybersecurity texts or cyber attack prediction based on our analyses.
%Surya - you need to present the summary of the findings as well. 
%tingmin - added a sentence to summarise two findings
%Surya - the article is well written, the analysis is good, but the paper fails to convince why do we need to do it? If I read this paper, what do I learn and what is the next piece of research. Sometimes the paper seems to reveal an obvious that we already know. Is it possible to bring the key messages upfront. I felt like they are now burried under the hood.  
%tingmin - added the key findings of each research question in Introduction.

%\section*{References}
\bibliographystyle{IEEEtran}
\bibliography{IEEEabrv,ref}

%\newpage

\begin{appendices}
\section{Optimal number selection of topics generated by LDA}
\label{a_LDA_optimal_number_selection}

\begin{figure}[htbp]%[htbp]
\centering
\includegraphics[width=0.95\linewidth]{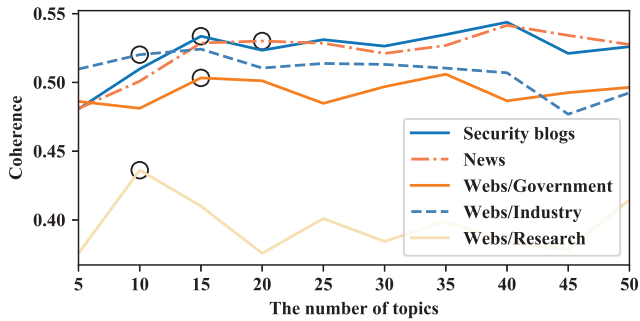}
\caption{The coherence of the models generated with the different number of topics (5, 10, ..., 50). The optimal number of topics is highlighted with a circle marker.}
\label{f_analysis_topic_selection}
\end{figure}

To identify the optimal number of the topics generated by LDA per dataset, we swept the 5 to 50 interval with steps of 5 and generated separate models accordingly. We used topic coherence  to measure the performance of the models \cite{roder2015exploring}. 
Fig.~\ref{f_analysis_topic_selection} depicts the results of the models.  The optimal one is marketed with a circle in each case. Our rule was to pick the model with the highest coherence. However, if the coherence did not increase much (difference $\leq$0.01) after the first peak, we kept the first highest value. For instance, the coherence for security blogs only rose by 0.01 from 15 to 40 topics, so we took 15 as the optimal topic number. If two models had similar coherence (e.g. news models with 15 and 20 topics), we manually compared the generated topics and selected more distinct one (e.g. news model with 20 topics). We manually read the generated topics and the articles per topics, and removed two topics in news model since both the topics and the articles are irrelevant to cybersecurity.

\section{The number of published articles in the last 20 years}
\label{a_no_published_articles}

\begin{figure}[htbp]%[htbp]
\centering
\includegraphics[width=1\linewidth]{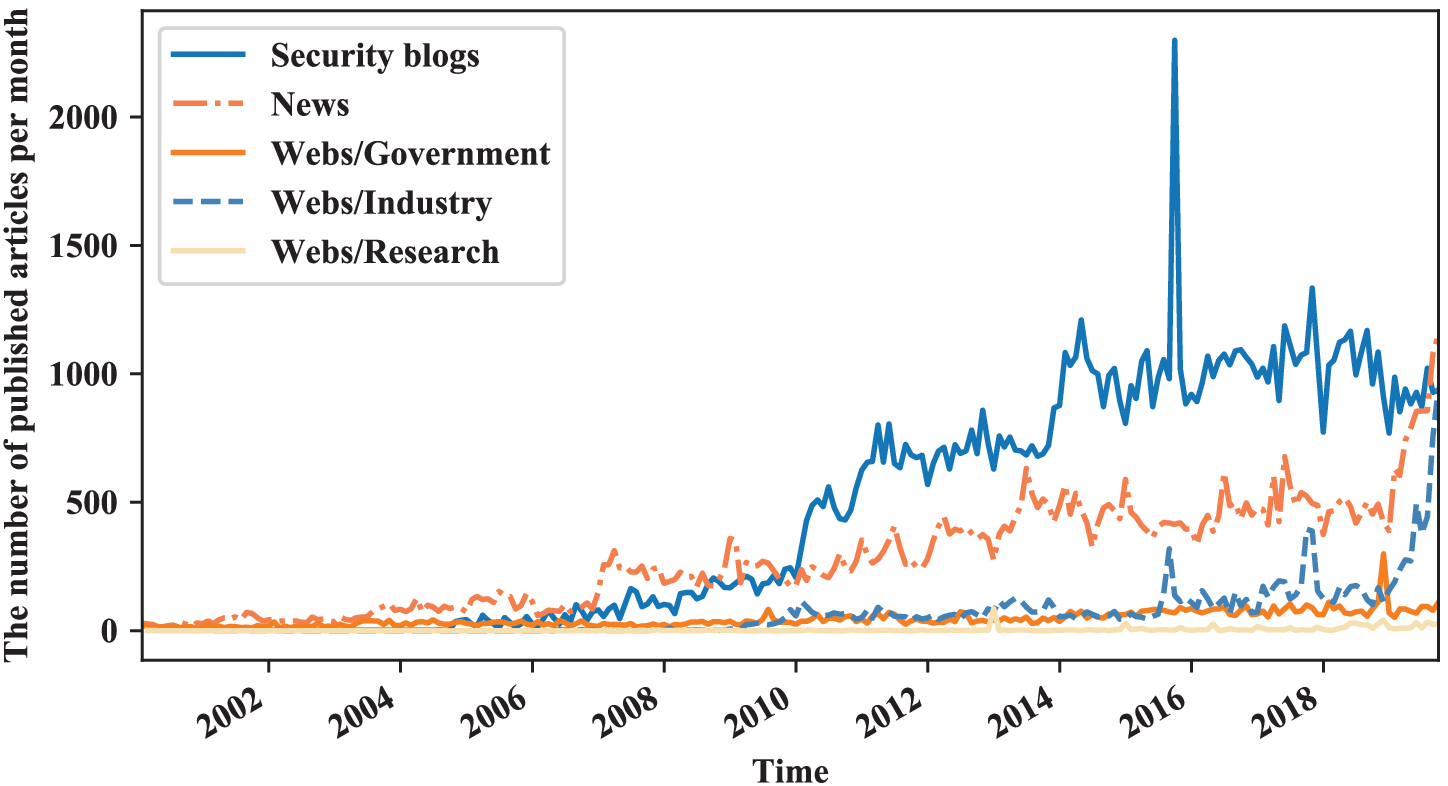}
\caption{The number of articles published from 2000 to now.}
\label{f_article_no_trend}
\end{figure}

Fig.~\ref{f_article_no_trend} shows the number of published articles per month in the five datasets in the last 20 years. Most of the cybersecurity texts came to surface after 2010, with a regular posting afterwards. We see that security blogs account for the majority of the online resources on security. With around half of the security blogs in post numbers, news volume shows a dramatic increase in the last year. Websites form a relative small portion, with a gentle growth over the time. In 2016 and 2018, two peaks can be spotted in the industrial websites curve. It has also jumped since last year. We observed that the number of published articles per month in the 2000s is far smaller than the number in the later ten years. Therefore, we mainly focused on the analysis of trends in the 2010s in our study.

\section{The number of terms after accommodating variants.}
\label{a_no_terms_clean}

%\begin{figure}[htbp]
\begin{figure}[htbp]
\centering
\includegraphics[width=1\linewidth]{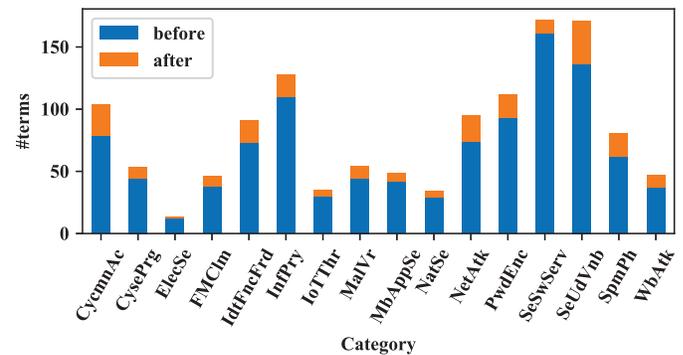}
\caption{The numbers of terms before and after identifying semantically and syntactically duplicated terms as term variants.}
\label{f_term_clean}
\end{figure}

\newpage

\section{Significant tests on the security categories}
\label{a_test_category_validation}
\begin{table}[htbp]
\centering
\caption{Three ${\chi}^2$ tests on the proportions of articles with each category as the most relevant across the LDA-generated topics, datasets and sources separately.}
%For each category, three ${\chi}^2$ tests are conducted to see whether the proportions are similar across the 68 LDA-generated topics, five datasets and three sources separately. %The $p$ values smaller than 0.01 indicate significant differences.
\label{t_p_security_categories}
\begin{tabular}{lll|ll|ll}
\hline
\multirow{2}{*}{Category} & \multicolumn{2}{c}{Across LDA topics} & \multicolumn{2}{c}{Across datasets}& \multicolumn{2}{c}{Across sources} \\ %\hline
\cline{2-7} %\cline{4-5}
    &  ${\chi}^2$ & $p$ &  ${\chi}^2$ & $p$&  ${\chi}^2$ & $p$ \\ \hline
CycmnAc & 7020 & <0.01 &978 &<0.01 & 768& <0.01 \\
CysePrg &8715 &<0.01 & 1528&<0.01 &1222 & <0.01 \\
ElecSe &6696 &<0.01 & 418&<0.01 &366 & <0.01 \\
FMClm &72679 &<0.01 &27129 & <0.01&7796 & <0.01 \\
IdtFncFrd &12393 & <0.01& 886&<0.01 &677 & <0.01 \\
InfPry &7216 &<0.01 & 566&<0.01 &554 & <0.01 \\
IoTThr &14526 &<0.01 &3241 &<0.01&3131 & <0.01 \\
MalVr &32297 &<0.01 &3230 & <0.01& 2758& <0.01 \\
MbAppSe &14383 & <0.01&1434 & <0.01& 1409& <0.01 \\
NatSe &26250 &<0.01 & 2688&<0.01 & 2335& <0.01 \\
NetAtk &14529 & <0.01& 1659&<0.01 &1253 & <0.01 \\
PwdEnc &17847 & <0.01& 2452& <0.01& 2060& <0.01 \\
SeSwServ &1075 &<0.01 &43 &<0.01 &25 & <0.01 \\
SeUdVnb &42059 & <0.01& 1992&<0.01 & 1773& <0.01 \\
SpmPh & 19368&<0.01 &1105 & <0.01&263 & <0.01\\
WbAtk &13433 &<0.01 & 3158&<0.01 & 1359 & <0.01 \\
\hline
\end{tabular}
\end{table}

%\newpage

\section{The popularity of our security categories}
\label{a_category_popularity}
\begin{figure}[htbp]
%\begin{figure}[t]
\centering
\includegraphics[width=1\linewidth]{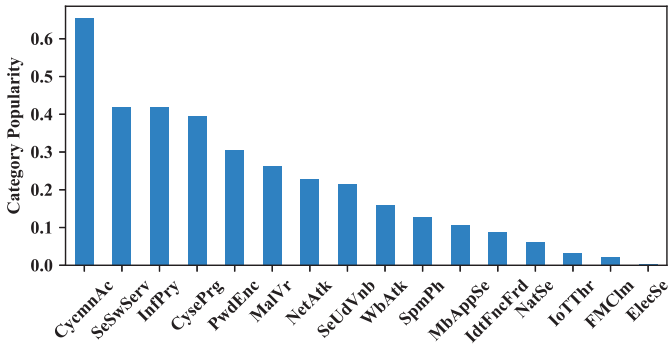}
\caption{The popularity of our security categories.}
\label{f_popularity_q2}
\end{figure}

\newpage
\section{Significance test on the trends of different sources}
\label{a_sig_test_sources}
\begin{table}[htbp]
\centering
\caption{Significance test on the trends of relative impact, including increasing ($\uparrow$), decreasing ($\downarrow$) and stable($\rightarrow$) trends ($p<0.05$).}
\label{t_p_rel_impact_q3_trend}
\begin{tabular}{llll}
\hline
Category & News & Security blogs & Webs \\ \hline
CycmnAc & $\uparrow$& $\uparrow$& $\downarrow$ \\
CysePrg & $\downarrow$ & $\uparrow$& $\downarrow$ \\
ElecSe & $\uparrow$ &$\uparrow$ & $\uparrow$ \\
FMClm & $\rightarrow$ & $\rightarrow$ & $\downarrow$ \\
IdtFncFrd &$\uparrow$ &$\uparrow$ & $\downarrow$ \\
InfPry & $\downarrow$& $\uparrow$& $\uparrow$ \\
IoTThr &$\downarrow$ & $\uparrow$& $\uparrow$ \\
MalVr & $\rightarrow$&$\uparrow$ & $\uparrow$ \\
MbAppSe & $\downarrow$& $\uparrow$& $\uparrow$ \\
NatSe & $\uparrow$& $\uparrow$& $\rightarrow$ \\
NetAtk &$\downarrow$ & $\uparrow$& $\uparrow$ \\
PwdEnc &$\uparrow$ & $\uparrow$& $\rightarrow$ \\
SeSwServ & $\rightarrow$&$\uparrow$ & $\uparrow$ \\
SeUdVnb &$\downarrow$ & $\uparrow$& $\rightarrow$ \\
SpmPh &$\uparrow$ &$\uparrow$ & $\uparrow$ \\
WbAtk &$\downarrow$ &$\uparrow$ & $\downarrow$ \\
\hline
\end{tabular}
\end{table}
%\section{B}
\end{appendices}

\end{document}